\documentclass[aps,pre,reprint,superscriptaddress,nofootinbib]{revtex4-2}

\usepackage{amsmath,amssymb,bm}
\usepackage{graphicx}
\usepackage{dcolumn}
\usepackage{hyperref}
\usepackage{xcolor}

\begin{document}

\title{Hidden Higher-Order Vulnerabilities in Simplicial Complexes Revealed by Fixed-Index Spectral Robustness}

\author{Kaiming Luo}
\email{kmluo24@m.fudan.edu.cn}
\affiliation{College of Future Information Technology, Fudan University, Shanghai 200438, China}

\date{Received 25 March 2026; revised 24 June 2026; accepted 25 August 2026; published 8 September 2026}

\begin{abstract}
Robustness of simplicial complexes under triangle deletion is often characterized by the instantaneous smallest positive eigenvalue of the Hodge $1$-Laplacian. We show that this observable can change spectral index when the kernel grows, and therefore it need not follow the same nonharmonic edge-space eigenmode along the deletion process. To remove this ambiguity, we define fixed-index spectral robustness by monitoring the eigenvalue whose index is selected by the first nonzero eigenvalue of the intact complex. The corresponding triangle sensitivity follows from first-order perturbation theory and identifies simplices that produce the steepest initial decrease of this monitored eigenvalue. Across synthetic and empirical clique complexes, a small fraction of triangles can drive the monitored eigenvalue to zero while the $1$-skeleton and graph-level robustness measures remain unchanged. These results show that higher-order robustness is a genuinely simplicial spectral property that cannot be inferred from graph-level connectivity alone.
\end{abstract}

\maketitle

\section{Introduction}

Higher-order networks and simplicial complexes provide a natural framework for describing interactions beyond pairwise links, with direct relevance to collective dynamics, spreading processes, synchronization, and signal processing on edge spaces \cite{Battiston2020,Salnikov2018,Bianconi2021,Battiston2021,Luo2024RemoteSync,Luo2026Hierarchical}. Empirical studies have shown that these descriptions capture meaningful organization in systems ranging from neural activity and brain connectomes to collaboration and group-interaction data \cite{Petri2014,Giusti2016,Sizemore2018,Benson2016}. At the same time, controlled higher-order ensembles and random simplicial-complex models have made it possible to isolate genuinely simplicial effects beyond any single dataset, including weighted growing constructions, Linial--Meshulam complexes, random clique complexes, and random geometric complexes \cite{Courtney2017,Linial2006,Meshulam2009,Bobrowski2018}. 

The Hodge $1$-Laplacian provides the natural spectral description of edge-space organization. Its nonzero spectrum separates exact, harmonic, and coexact components and controls relaxation rates for linear edge-space dynamics \cite{Eckmann1945,Horak2013,Jiang2011,Lim2020,Schaub2020,Barbarossa2020}. This makes Hodge spectra natural candidates for higher-order robustness observables. At the same time, the spectrum depends on the coface structure of the complex, so a perturbation that leaves the graph unchanged can still modify the higher-order operator.

In this setting, dismantling problems analogous to percolation or targeted attacks must also be reconsidered \cite{Callaway2000,Cohen2001,Albert2000,Morone2015,Braunstein2016}, because system function is not solely determined by the node--edge skeleton. In particular, simplices can be removed while leaving all nodes and edges strictly unchanged; graph-level connectivity and standard robustness indicators therefore remain identical, yet the higher-order functions supported by those simplices may weaken or collapse. Deleting simplices can thus leave the graph unchanged while destroying function, indicating that graph robustness and higher-order functional robustness can be fundamentally distinct.

The relevant question is therefore not whether the graph remains connected, but whether a specific higher-order functional channel present in the intact complex survives simplex deletion. Higher-order random walks, spectral inequalities, and edge-space diffusion formalisms all point to Hodge-type operators as the correct carriers of topology-aware dynamics on simplicial complexes \cite{Mukherjee2016,Parzanchevski2017,Steenbergen2014}. However, the instantaneous smallest positive eigenvalue is not stable under simplex removal. As simplices are deleted, the harmonic subspace may expand, causing the observable to switch between spectral branches. Branch switching changes the identity of the monitored mode: the observable no longer tracks the same functional channel, but instead reports a different mode that has become minimal.

The problem is therefore definitional before it is algorithmic. If robustness is intended to quantify the fate of a specific functional channel, then the observable must remain tied to that channel throughout the perturbation. This requires defining robustness on a fixed spectral branch, rather than on the branch that is instantaneously minimal. A natural choice is to fix the first nonharmonic branch of the intact complex and track it continuously during simplex deletion. Only under this branch-consistent definition does functional weakening acquire a well-defined physical meaning.

In this work, we adopt this perspective and redefine higher-order functional robustness at the level of the observable. Within this fixed-index framework, Mode Sensitivity (MS) arises naturally as the simplex-level local response of the tracked branch. MS is not an independently introduced heuristic ranking, but the first-order perturbative sensitivity of the fixed spectral mode to simplex removal \cite{Kato1995,Stewart1990}. The induced removal direction therefore follows directly from the definition of the observable, rather than preceding it.

This redefinition leads to a clear physical consequence. On both synthetic clique complexes and empirical networks, removing only a small fraction of simplices is sufficient to drive the tracked nonharmonic mode to collapse, while the graph skeleton remains strictly unchanged. Graph-level robustness indicators are therefore exactly constant throughout the process, and the observed failure is purely higher-order. The collapse does not originate from structural damage to the graph, but from the destruction of the functional channel supported by the simplicial structure.

This strict separation reveals a hidden higher-order vulnerability. A system may appear fully intact at the graph level, with all nodes and edges preserved, yet its essential higher-order functional mode has already vanished. Conventional graph-based measures fail to detect this failure not because they are insufficiently sensitive, but because they monitor an object that does not change. Only by defining robustness on a fixed spectral branch, and tracking the same mode throughout, can functional collapse be correctly identified as the disappearance of a single physical channel rather than a relabeling of modes.

The remainder of this paper develops this logic. We first establish a branch-consistent definition of higher-order functional robustness and demonstrate the failure of the conventional observable under simplex deletion. We then derive MS from perturbation theory, linking the observable to a local dismantling mechanism. Finally, we provide numerical evidence that rapid functional collapse occurs even when the graph remains unchanged. The central conclusion is that higher-order robustness is systematically mischaracterized unless it is defined on a fixed spectral branch.

\section{Fixed-Index Spectral Robustness}
\label{sec:Def}

We consider a general weighted, possibly directed graph 
\(
G=(V,E,W),
\)
where \(w_{ij}\ge 0\) denotes the coupling strength from node \(j\) to \(i\). 
The associated clique complex is restricted to triangles, with edges and triangles oriented consistently. 
Let \(B_1\) and \(B_2\) denote the node-edge and edge-triangle incidence matrices, respectively. 
For weighted graphs, the Hodge $1$-Laplacian is defined as
\begin{equation}
    L_1 = B_1^\top W_1 B_1 + B_2 W_2 B_2^\top,
\end{equation}
where \(W_1 = \mathrm{diag}(w_e)\) encodes edge weights, and \(W_2\) encodes triangle weights (either unit or edge-weight-based). 
For directed networks, one can take the Hermitian part \(
L_1 = \frac{L_1 + L_1^\dagger}{2}
\)  to ensure real eigenvalues. 
The spectrum of \(L_1\)  then identifies the slowest nontrivial edge-space response channels.

Eigenvalues are ordered as
\begin{equation}
    0 = \lambda_1 \le \lambda_2 \le \cdots \le \lambda_m,
\end{equation}
with the zero modes corresponding to harmonic edge flows. 
Denoting the harmonic subspace at removal fraction \(f\) as 
\(
\mathcal{H}(f) = \ker L_1(f),
\)
the Hodge decomposition
\begin{equation}
    \mathbb{R}^{m} = \operatorname{im}(B_1^\top) \oplus \ker L_1 \oplus \operatorname{im}(B_2)
\end{equation}
separates exact, harmonic, and coexact contributions \cite{Eckmann1945,Jiang2011,Lim2020}. 
Physically, the first nonharmonic branch corresponds to the softest edge-space excitation, and its evolution under simplex deletion reveals the fragility of higher-order structure.

Let $k_* = \dim\ker L_1(0)+1$ denote the index of the first nonharmonic eigenvalue in the intact complex. Removing a fraction $f$ of triangles modifies the Laplacian as
\begin{equation}
    L_1(f) = L_1(0) - \sum_{\tau\in \mathcal{R}(f)} W_\tau^{1/2} b_\tau b_\tau^\top W_\tau^{1/2},
\end{equation}
where $b_\tau$ is the oriented boundary vector and $W_\tau$ its weight. Each term is negative semidefinite, ensuring the Loewner order
\begin{equation}
    L_1(f_2) \preceq L_1(f_1), \quad f_2>f_1,
\end{equation}
so that any fixed-index eigenvalue decreases monotonically under triangle removal. This observation underlies the branch-consistent robustness observable
\begin{equation}
    \widetilde{\mu}_1(f) = \lambda_{k_*}[L_1(f)],
\end{equation}
whose associated eigenvector $\widetilde{u}(f)$ preserves the identity of the monitored functional branch. Instantaneous minima
\begin{equation}
    \mu_+^{\mathrm{inst}}(f) = \lambda_{\dim\ker L_1(f)+1}[L_1(f)]
\end{equation}
suffer branch-switching ambiguities, potentially misrepresenting mode weakening. In this formulation, $\widetilde{\mu}_1(f)$ provides a rigorous anchor for higher-order functional analysis, independent of deletion strategy.

To quantify the decay of the tracked mode, we define the normalized robustness
\begin{equation}
    R_{\mathrm{HO}}(f) = \frac{\widetilde{\mu}_1(f)}{\widetilde{\mu}_1(0)},
\end{equation}
which directly reflects the weakening of the monitored functional branch.

\begin{figure*}[t]
    \includegraphics[width=0.9\linewidth]{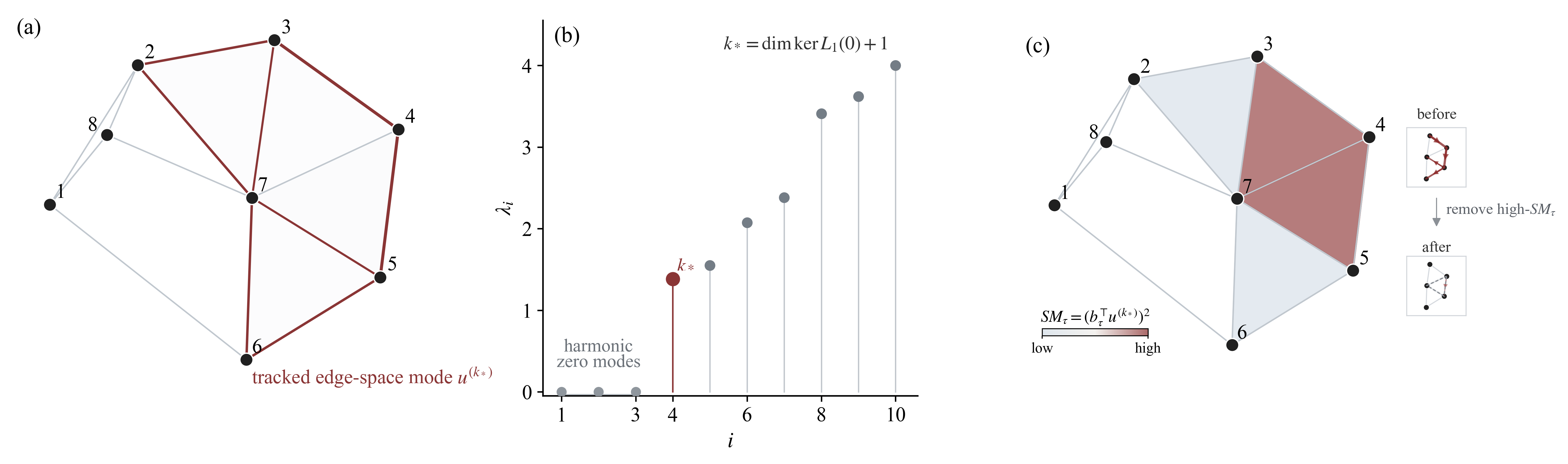}
    \caption{Branch-consistent functional robustness in weighted and directed clique complexes. High-sensitivity triangles are colored red; tracked mode edges are indicated in red arrows; key triangle outlined in yellow. The figure serves as an illustration of the logic: defining the tracked mode, freezing its spectral index, and scoring simplices by Mode Sensitivity. Insets show suppression of the same branch after triangle removal.}
    \label{fig:model}
\end{figure*}

The branch has a direct dynamical interpretation: for edge-space relaxation
\begin{equation}
    \dot{\bm{y}} = - L_1(f) \bm{y},
\end{equation}
the tracked modal amplitude decays exponentially
\begin{equation}
    \widetilde{u}(f)^\top \bm{y}(t) = (\widetilde{u}(f)^\top \bm{y}(0)) e^{-\widetilde{\mu}_1(f) t},
\end{equation}
with energy decay
\begin{equation}
    \|\bm{y}(t)\|_2^2 \approx C(f) e^{-2 \widetilde{\mu}_1(f) t}, \quad C(f)=|\widetilde{u}(f)^\top \bm{y}(0)|^2,
\end{equation}
defining a physically interpretable timescale
\begin{equation}
    t_\varepsilon \approx \frac{1}{2 \widetilde{\mu}_1(f)} \log\left(\frac{C(f)}{\varepsilon}\right),
\end{equation}
which can be directly compared across networks of varying weight and directionality. This timescale illustrates that even small-weight triangles, if highly aligned with the tracked mode, can dominate higher-order functional decay.

The effect of removing a single triangle $\tau$ can be examined through the one-parameter perturbation
\begin{equation}
    L_1(s) = L_1(0) - s\, W_\tau^{1/2} b_\tau b_\tau^\top W_\tau^{1/2}, 
    \qquad 0 \le s \le 1,
\end{equation}
which yields the first-order eigenvalue response
\begin{equation}
    \left.\frac{d\widetilde{\mu}_1}{ds}\right|_{s=0}
    = -\bigl(b_\tau^\top W_\tau^{1/2} u\bigr)^2
    = -MS_\tau .
\end{equation}
This defines the Mode Sensitivity (MS) of triangle $\tau$. The quantity $MS_\tau$ is therefore not an independently introduced heuristic, but the first-order downward response of the tracked nonharmonic branch to simplex removal. Physically, it measures how efficiently a given triangle suppresses the monitored functional channel.

A global interpretation follows immediately:
\begin{equation}
    \sum_{\tau\in \mathcal{T}} MS_\tau
    = u^\top B_2 W_2 B_2^\top u
    = \|W_2^{1/2} B_2^\top u\|_2^2,
\end{equation}
so the total sensitivity mass is exactly the coexact energy carried by the tracked mode. If this sensitivity field is strongly concentrated on a small subset of simplices, then removing only a small fraction of triangles can already induce rapid functional collapse, independently of local triangle density or node degree, which only probe geometric prominence rather than modal support.

This interpretation directly links the sensitivity field to the decay of the tracked branch.  Let $z_\tau\in\{0,1\}$ indicate whether triangle $\tau$ is removed. Linearizing the tracked eigenvalue around the intact complex gives
\begin{equation}
    \widetilde{\mu}_1(\mathbf{z})
    \approx
    \widetilde{\mu}_1(0)
    -
    \sum_{\tau\in\mathcal{T}} z_\tau MS_\tau .
\end{equation}
To leading order, removing the simplices with the largest $MS_\tau$ therefore produces the steepest descent of the tracked branch. The static MS ranking should thus be understood as the first-order descent rule induced by the branch-consistent observable, rather than as an independently postulated attack strategy.

Branch-consistent functional robustness therefore reveals that the fragility of the tracked mode is governed by its alignment with a small subset of critical triangles, rather than by geometric density or combinatorial prominence alone. The field $MS_\tau$ identifies which simplices control higher-order functional collapse and shows that vulnerability is a property of the mode itself.

This conceptual hierarchy is illustrated in Fig.~\ref{fig:model}. The intact clique complex displays the tracked edge-space mode $u(k_*)$ on the 1-skeleton, while a representative high-sensitivity triangle is highlighted. The initial $L_1$ spectrum marks the fixed index $k_*$ that anchors the branch-consistent observable. The induced $MS_\tau$ field then maps the spatial support of the tracked mode over simplices, and the before/after insets show that removing a high-$MS_\tau$ triangle suppresses the same branch, making explicit the link between mode alignment and higher-order collapse.

\section{Results}

We begin by fixing the branch-consistent observable defined in Sec.~\ref{sec:Def} to ensure that functional weakening is tracked on a single physical mode throughout simplex removal. This choice isolates the effect of triangle deletion from structural changes to the underlying graph skeleton and provides a physically meaningful measure of higher-order robustness. Using this observable, we then examine (i) the fraction of triangles required to collapse the tracked mode, (ii) the exact separation between higher-order functional collapse and graph-level connectivity, (iii) the localization of critical simplices that mediate the collapse, and (iv) the transfer of these patterns to empirical networks as well as their dynamical consequences. Comparisons with local, graph-based, and random heuristics are included only as diagnostic surrogates to highlight that the observed collapse arises from the intrinsic sensitivity of the branch-consistent mode rather than from the choice of ranking protocol.

\begin{figure}[t]
    \centering
    \includegraphics[width=1\linewidth]{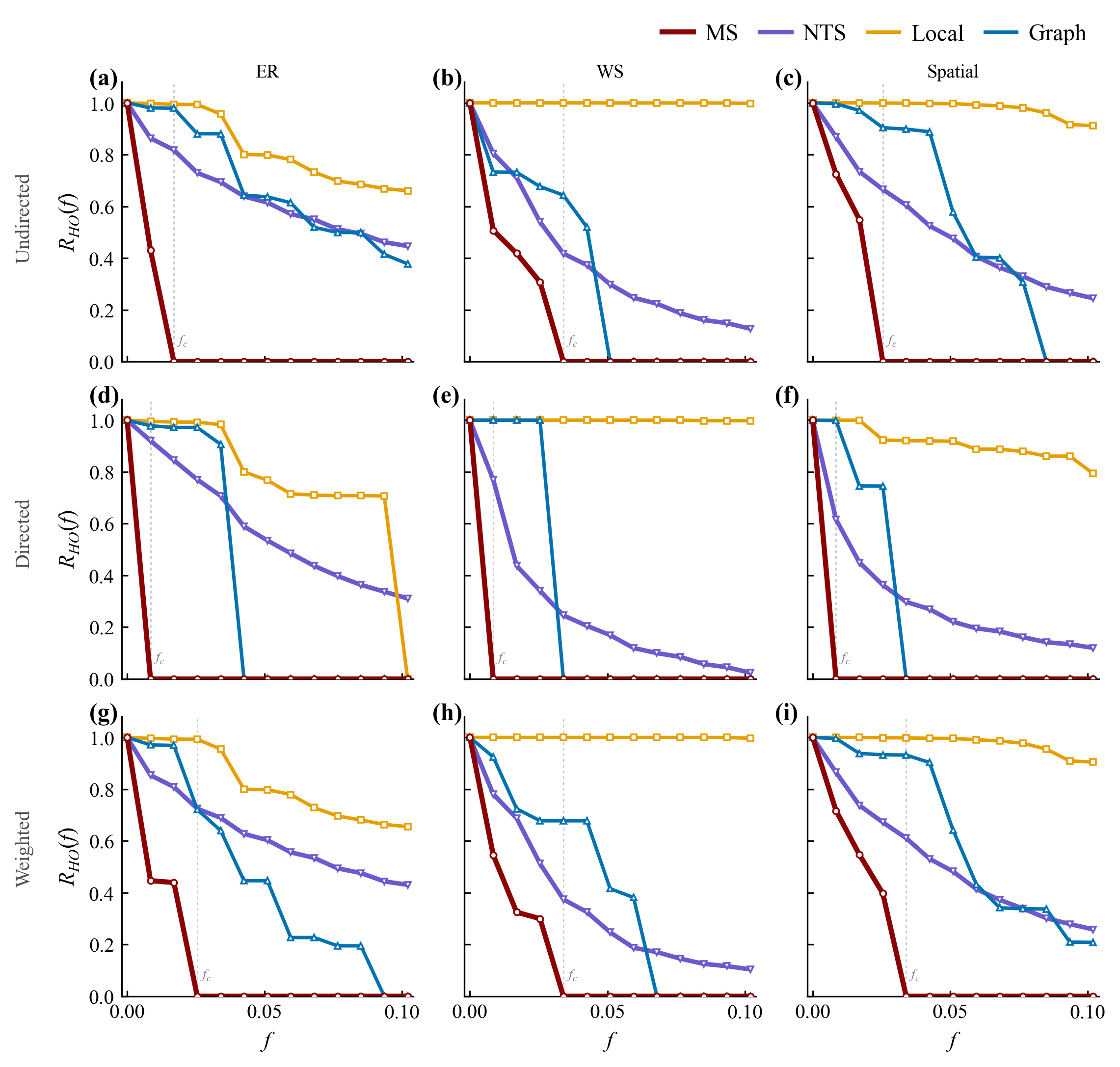}
\caption{
Functional dismantling on synthetic clique complexes probed by the branch-consistent observable 
$R_{\mathrm{HO}}(f)=\widetilde{\mu}_1(f)/\widetilde{\mu}_1(0)$.
Columns correspond to Erd\H{o}s--R\'enyi, Watts--Strogatz, and spatial random ensembles; 
rows show undirected, directed, and weighted cases.
Removing triangles ranked by mode sensitivity (MS) collapses the tracked higher-order mode 
after removing only a tiny fraction of simplices, in sharp contrast to both the non-tracked spectral baseline (NTS) 
and graph-based heuristics (Local, Graph), which display gradual decay.
The dashed line indicates the critical fraction $f_c$ where the tracked mode vanishes.
}
    \label{fig:synthetic}
\end{figure}

\subsection{Rapid Collapse of the Tracked Higher-Order Mode}

The central numerical observation is that the tracked higher-order mode is controlled by a very small subset of simplices. Once triangles are removed in the order prescribed by the branch-consistent sensitivity field, the normalized robustness
\(
R_{\mathrm{HO}}(f)=\widetilde{\mu}_1(f)/\widetilde{\mu}_1(0)
\)
drops much faster than under any of the three surrogate rankings shown in Fig.~\ref{fig:synthetic}. 

The effect is systematic across the three benchmark ensembles and therefore does not depend on a single geometric organization of triangles. In the Erd\H{o}s-R\'enyi complex, the tracked mode is extinguished already at \(f=0.015\), while the local and graph baselines still retain \(78\%\) and \(61\%\) of the initial robustness at \(f=0.06\). In the Watts-Strogatz complex, collapse occurs at \(f=0.035\) under MS, earlier than under both local and graph rankings. In the spatial benchmark, the same branch-consistent strategy again reaches zero first, at \(f=0.025\), while the three baselines all leave a substantial residual value over the same deletion window. 

The common message is that higher-order functional collapse is highly concentrated: only a very small removal budget is needed, provided that the removed simplices are chosen according to the support of the tracked nonharmonic mode rather than according to generic structural prominence.

To interpret the collapse curves in Fig.~\ref{fig:synthetic}, we compare the MS ranking with three diagnostic baselines. The first baseline is the local overlap score
\begin{equation}
S_{\tau}^{(\mathrm{local})}
=
\sum_{e\subset\tau}(n_e-1),
\end{equation}
where \(n_e\) is the number of triangles incident on edge \(e\). This score tests a purely local hypothesis: if higher-order vulnerability were mainly controlled by triangle redundancy or by residence in triangle-rich regions, then removing simplices with the largest overlap should already approximate the optimal attack.

The second baseline is the graph degree-sum score
\begin{equation}
S_{\tau}^{(\mathrm{graph})}
=
k_a+k_b+k_c,
\qquad \tau=(a,b,c),
\end{equation}
with \(k_a\), \(k_b\), and \(k_c\) the degrees of the three vertices of triangle \(\tau\). This baseline probes a different surrogate mechanism. If functionally critical simplices were simply attached to graph-prominent nodes, then a ranking inherited from the graph skeleton should already identify them, even though the attack itself acts only on 2-simplices.

The third baseline is the non-tracked spectral baseline (NTS). In contrast to MS, it does not follow the fixed branch \(k_*\) inherited from the intact complex, but instead ranks triangles through the instantaneous smallest positive eigenvalue
\begin{equation}
\mu_{+}^{\mathrm{inst}}(f)
=
\lambda_{\dim\ker L_1(f)+1}\!\left[L_1(f)\right].
\end{equation}
Operationally, NTS asks whether a spectral strategy that ignores branch identity is already sufficient. This is the most stringent baseline in conceptual terms, because it uses spectral information from the Hodge operator itself, but does so with the wrong observable: the monitored quantity is always the currently smallest positive eigenvalue, regardless of whether it still belongs to the original functional channel.

These three baselines are not redundant. The local score isolates the role of simplex crowding, the graph score isolates the role of node prominence inherited from the 1-skeleton, and NTS isolates the role of spectral information without branch consistency. Taken together, they separate three plausible but distinct explanations for rapid collapse. The fact that MS outperforms all three shows that the vulnerable simplices are not selected by local redundancy, are not reducible to graph-central locations, and are not recoverable from a reindexed spectral observable that changes identity during deletion.

The difference between MS and the two structural surrogates, Local and Graph, is already physically informative. In clique complexes these two baselines often behave similarly because both favor dense triangle-rich regions: large endpoint degrees tend to correlate with large overlap counts, so both scores are biased toward simplices embedded in locally crowded parts of the complex. Their slower decay curves in Fig.~\ref{fig:synthetic} therefore show that the collapse mechanism is not governed by density alone. What matters is not how many neighboring triangles a simplex has, nor how highly connected its vertices are in the graph, but how strongly that simplex loads onto the tracked nonharmonic mode. The MS ranking is effective precisely because it resolves this mode support at the level of the Hodge operator.

The difference between MS and NTS is more fundamental still, because it concerns the definition of the observable itself. Under simplex deletion, the tracked eigenvalue
\(
\widetilde{\mu}_1(f)
\)
is monotone nonincreasing by construction, and therefore faithfully quantifies the weakening of one fixed physical channel. By contrast, the instantaneous quantity
\(
\mu_{+}^{\mathrm{inst}}(f)
\)
need not describe the same channel at different values of \(f\). When the harmonic subspace grows, the positive spectrum is reindexed, and the smallest positive eigenvalue may jump from the original nonharmonic branch to a different one. NTS therefore does not merely provide a weaker ranking; it optimizes a quantity whose physical identity is not preserved along the dismantling trajectory.

\begin{figure}[t]
    \centering
    \includegraphics[width=0.9\linewidth]{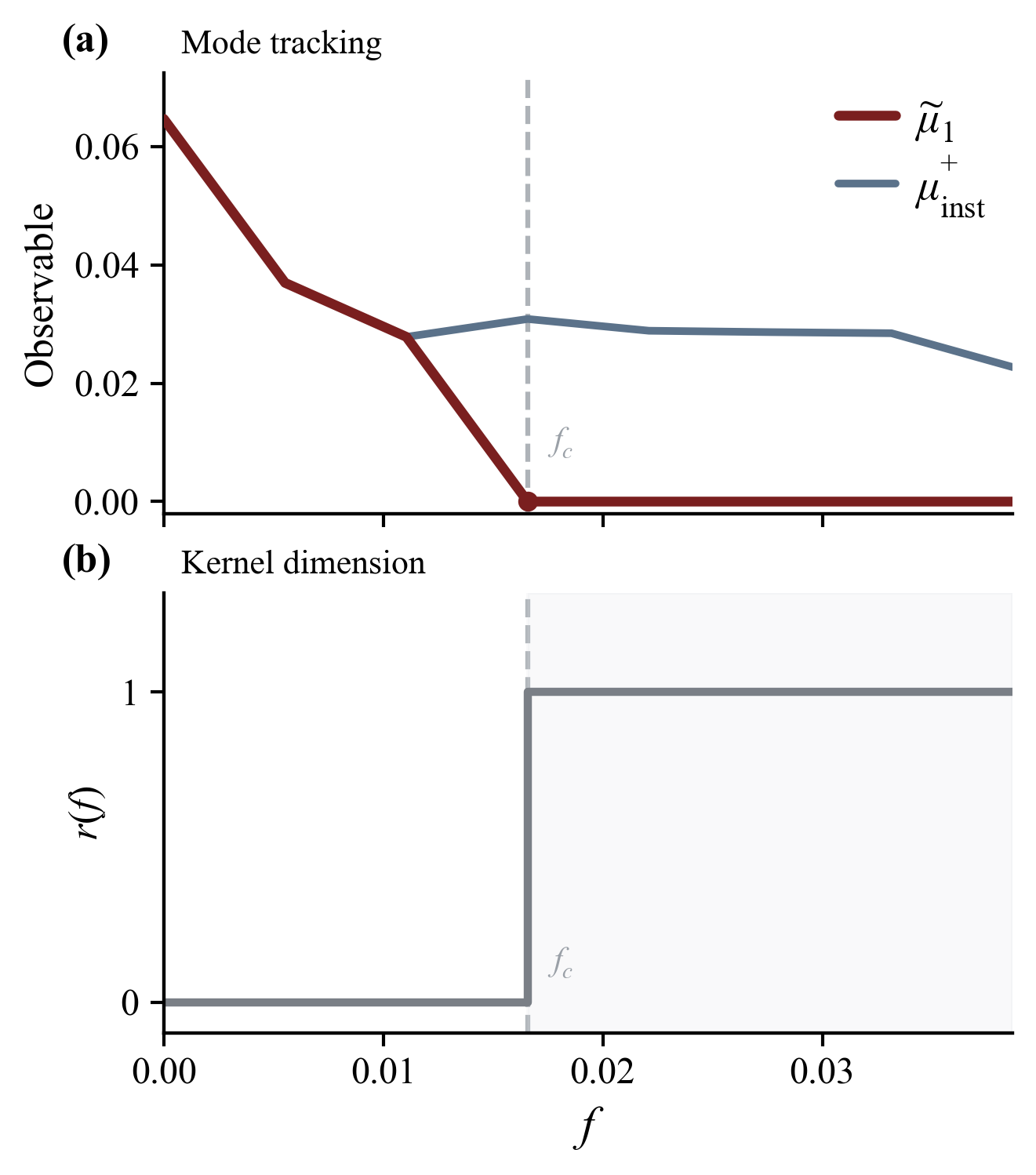}
    \caption{
    Failure of the instantaneous smallest positive eigenvalue as a robustness observable.
    (a) The branch-consistent eigenvalue \(\widetilde{\mu}_1(f)\) (red) vanishes at the critical fraction \(f_c\), signaling collapse of the tracked higher-order mode, whereas the instantaneous smallest positive eigenvalue \(\mu_{+}^{\mathrm{inst}}(f)\) (blue) remains finite because it switches to a different branch after the tracked mode enters the kernel.
    (b) The kernel dimension increases at \(f_c\), identifying the emergence of an additional harmonic mode.
    The coincidence of tracked-mode collapse and kernel growth shows that \(\mu_{+}^{\mathrm{inst}}(f)\) does not monitor a fixed physical degree of freedom and therefore cannot serve as a consistent functional robustness observable.
    }
    \label{fig:ER_branch_switch}
\end{figure}

This point is resolved directly in Fig.~\ref{fig:ER_branch_switch}. The branch-consistent eigenvalue \(\widetilde{\mu}_1(f)\) decreases to zero at the critical fraction \(f_c\), marking the collapse of the tracked higher-order mode. The corresponding event is not a smooth attenuation of a generic positive eigenvalue, but a genuine transfer of that mode into the harmonic sector. At the same removal fraction, the kernel dimension increases discontinuously, showing that a new harmonic mode has been created. The vanishing of \(\widetilde{\mu}_1(f)\) and the jump in \(\dim\ker L_1(f)\) are therefore two sides of the same spectral transition.

The instantaneous smallest positive eigenvalue behaves differently. Instead of collapsing at \(f_c\), it remains finite because, once the tracked branch reaches zero and enters the kernel, the observable switches to the next positive branch. The apparent persistence of \(\mu_{+}^{\mathrm{inst}}(f)\) beyond \(f_c\) is thus not evidence of residual robustness of the original mode. It is a bookkeeping artifact caused by branch switching. In physical terms, the observable has changed the degree of freedom it monitors. This is exactly why NTS underperforms: even though it is spectral, it ranks simplices with respect to a quantity that ceases to represent the same functional channel throughout the attack.

The combined evidence from Figs.~\ref{fig:synthetic} and \ref{fig:ER_branch_switch} therefore supports a sharper conclusion than a simple benchmarking statement. Rapid functional collapse is possible because the tracked nonharmonic mode is strongly concentrated on a small simplex set, and this concentration can only be exposed by an observable that preserves branch identity. Local and graph surrogates fail because they probe generic structural prominence rather than tracked-mode support. NTS fails for a different reason: it remains spectral but abandons the physical identity of the mode when the spectrum is reindexed. The superiority of MS is thus not merely algorithmic. It reflects that MS is the only ranking among the four curves that is derived from a well-defined higher-order functional observable.

\subsection{Decoupling of Functional and Structural Robustness}
\begin{figure}[t]
    \includegraphics[width=1\linewidth]{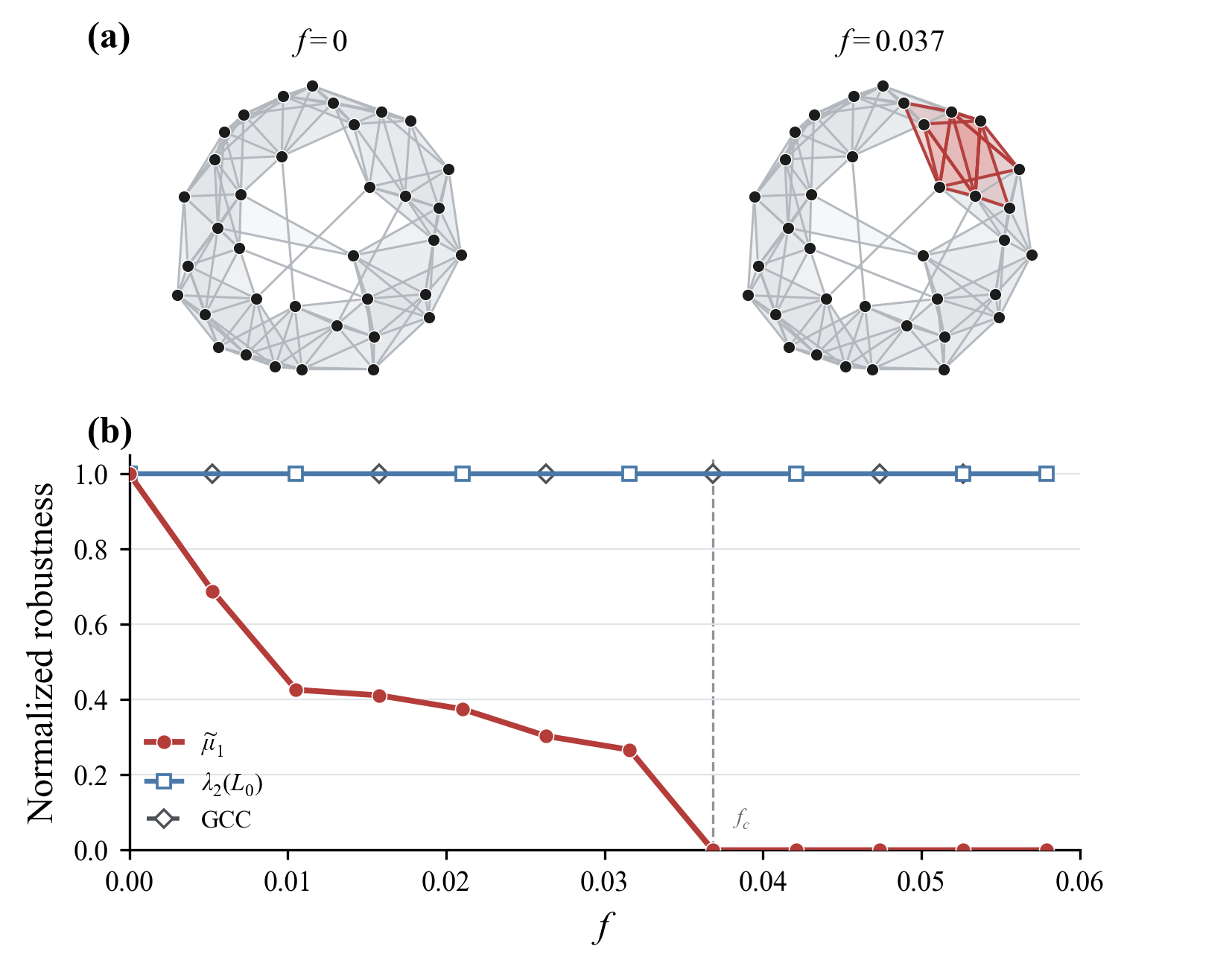}
    \caption{Higher-order functional collapse missed by graph benchmarks. (a) Watts-Strogatz clique complex at $f=0$ and at the realized collapse threshold $f_c\approx 0.037$, with the actually removed simplices shown as translucent red triangular regions. The node-edge skeleton is visually unchanged because only $2$-simplices are deleted. (b) The branch-consistent robustness $\widetilde{\mu}_1(f)$ collapses to zero at $f_c$, whereas the normalized graph observables normalized $\lambda_2(L_0)$ and GCC remain identically equal to $1$ throughout, since triangle deletion leaves the $1$-skeleton exactly invariant. The dashed vertical line marks the mode-collapse threshold.}
    \label{fig:separation}
\end{figure}

Because only triangles are removed, higher-order functional weakening can be isolated exactly from graph-level connectivity. In that setting, Fig.~\ref{fig:separation} serves as a validation of the observable itself, not merely another performance comparison. Using the Watts-Strogatz benchmark, we compare the tracked mode against two graph-level robustness indicators. While the higher-order robustness $\widetilde{\mu}_1(f)$ decays steadily and vanishes at the realized collapse threshold $f_c\approx 0.037$, both $\lambda_2(L_0)$ and the GCC remain fixed at their unperturbed values throughout the protocol because the graph skeleton is unchanged. In other words, the complex loses its higher-order functional mode before any graph-theoretic notion of connectivity changes at all.

This separation is important conceptually because it is exact, not approximate, in the present setting. Since $L_0(f)=L_0(0)$ by construction, every graph-level spectral change is ruled out a priori. The higher-order collapse shown in Fig.~\ref{fig:separation} is therefore not an artifact of graph fragmentation or edge pruning; it is a genuinely higher-order effect generated by deleting a few carefully chosen simplices while leaving the ordinary network untouched.

This makes Fig.~\ref{fig:separation} the sharpest qualitative result of the paper. It shows that higher-order functional robustness should be treated as a distinct observable, not as a by-product of graph robustness. A pairwise description can certify perfect structural integrity at the same time that the tracked nonharmonic mode has completely collapsed.

Panel (a) makes this mismatch visually concrete. After the top-ranked triangles are removed, the drawing still looks almost unchanged at the level of nodes and edges: the same skeleton persists, and no graph-level observer would infer major structural damage from the picture alone. What has changed is the pattern of filled $2$-simplices. The removed triangles are sparse enough that the intervention appears mild in ordinary network terms, yet panel (b) shows that this sparse coface surgery is sufficient to drive the tracked mode to zero. The force of the figure lies precisely in that contrast between visual graph-level continuity and spectral higher-order collapse.

The flat graph-level curves in panel (b) should therefore be read literally, not impressionistically. They are not merely ``small'' compared with the higher-order curve; they are constrained to be unchanged by the construction of the protocol. The entire vertical separation between $\widetilde{\mu}_1(f)$ and the graph observables is thus attributable to the higher-order term of the Hodge operator. This makes Fig.~\ref{fig:separation} more than an illustrative example. It functions as a controlled physical argument that branch-consistent higher-order robustness is measuring a phenomenon that pairwise structure cannot see.

\begin{figure}[t]
    \includegraphics[width=1\linewidth]{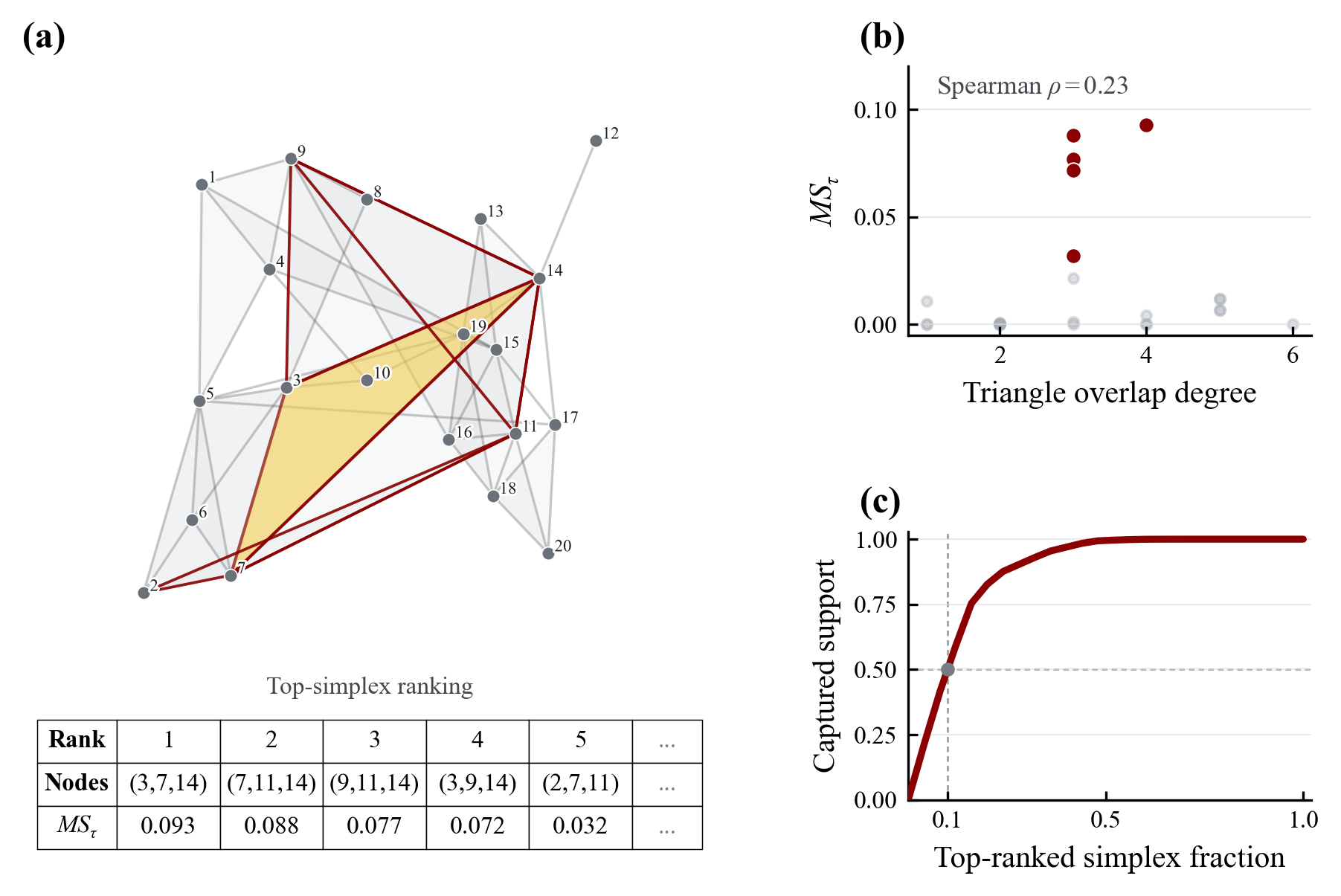}
    \caption{Localization and concentration of critical simplices in a modular bridge complex. (a) Triangle support map together with the top-simplex ranking table. The highest-ranked simplices concentrate on the intercommunity bridge rather than inside dense modules. (b) Sensitivity $MS_\tau$ versus triangle-overlap degree, showing weak correlation ($\rho=0.23$). (c) Cumulative captured support versus ranked simplex fraction; a small fraction of simplices carries a large portion of the total support.}
    \label{fig:critical}
\end{figure}

\begin{figure}[t]
    \includegraphics[width=1\linewidth]{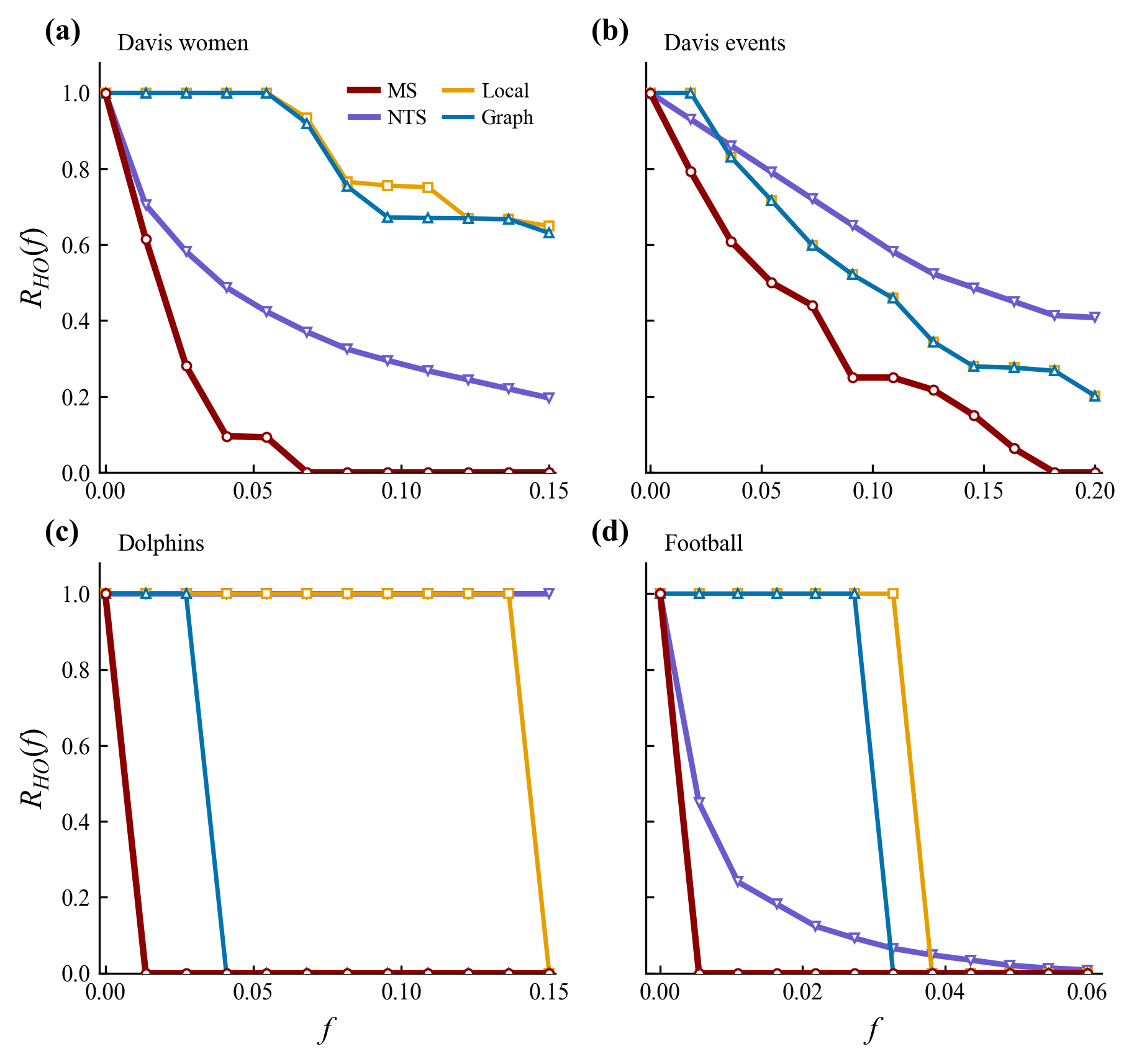}
    \caption{Functional dismantling on four representative empirical graphs: (a) Davis women, (b) Davis events, (c) Dolphins, and (d) Football. Branch-consistent sensitivity (MS) produces the fastest collapse of the tracked higher-order mode in all cases. The non-tracked spectral baseline (NTS) does not follow the same spectral branch, while local and graph heuristics are systematically less effective.}
    \label{fig:real}
\end{figure}

\subsection{Localization and Universality of Critical Simplices}

The efficiency of sensitivity-guided dismantling is controlled by where the tracked higher-order mode is supported in simplex space. The crucial point is that the simplices most responsible for functional collapse need not lie in the most triangle-rich parts of the complex. Instead, the tracked mode can concentrate on a sparse bridge carrier that mediates communication between larger mesoscopic regions. In that regime, simplex importance is determined by modal support on the tracked branch rather than by local redundancy.

The dominant simplices are localized along the intercommunity interface, not inside the dense modules, as shown in Fig.~\ref{fig:critical}. The support map in Fig.~\ref{fig:critical}(a) shows that the highest-ranked triangles align along a narrow conduit connecting the two sides of the complex, indicating that the tracked edge-space mode is funneled through a bridge-like carrier. The associated ranking confirms that these simplices are selected because they bear the largest projection of the tracked mode, not because they maximize triangle abundance. In physical terms, they carry the coexact response whose suppression drives the observed functional collapse.

The two statistical diagnostics further clarify why this localization is so destructive. As shown in Fig.~\ref{fig:critical}(b), sensitivity and triangle-overlap degree are only weakly correlated, with Spearman coefficient $\rho=0.23$. Sensitivity is therefore not a surrogate for local density, but identifies the structural sector on which the tracked branch is actually supported. In addition, the cumulative support curve in Fig.~\ref{fig:critical}(c) shows strong concentration: a small fraction of ranked simplices already captures a disproportionately large share of the total sensitivity weight. The rapid initial rise followed by a long tail indicates a highly uneven support landscape.

This support concentration explains why only a small number of simplices is sufficient to induce collapse. Because the cumulative contribution
\begin{equation}
    \sum_{\tau\in\mathcal{R}(f)} MS_\tau
\end{equation}
grows rapidly at small ranked fraction $f$, the tracked branch can be strongly suppressed before any extensive removal is needed. The mechanism is therefore geometric and spectral at the same time: the mode is funneled through a sparse bridge carrier, and its weight is strongly concentrated on a few critical simplices. Once those simplices are removed, the functional branch collapses quickly even though most of the complex remains intact. This also explains why density-based local and graph-level heuristics systematically underperform: they preferentially target simplex-rich cores, whereas the tracked mode is often controlled by a sparse support set outside those cores.

The same support-driven mechanism persists beyond stylized constructions. This phenomenon survives in empirical graphs, as shown in Fig.~\ref{fig:real}. Across the women and event projections of the Davis Southern Women network \cite{Davis1941}, as well as the Dolphins and Football networks\cite{nr}, sensitivity-guided deletion again produces the fastest collapse of the tracked higher-order mode. Although these systems differ substantially in size, triangle organization, and community structure, the same branch-consistent observable reveals a small vulnerable simplex sector that governs the higher-order response.

The empirical comparison also confirms that the non-tracked spectral baseline remains inadequate. NTS ranks simplices using the instantaneous smallest positive eigenvector after each deletion step, so it does not preserve the identity of the monitored mode when branch switching occurs. By contrast, MS is tied to the branch fixed in the intact complex and therefore continues to target the simplices that most efficiently suppress that same functional mode. This distinction is consistently observed across all empirical cases: once branch consistency is enforced, MS isolates a more destructive removal set than NTS, while local and graph heuristics remain less effective.

What varies across empirical graphs is not the mechanism itself but the strength of localization. In Dolphins and Football, the tracked mode is sharply concentrated, so removing a very small simplex fraction already produces near-instant collapse. In the Davis events projection, the decay is broader, indicating that support is more distributed across simplices. The Davis women projection lies between these extremes. The localization strength therefore changes from network to network, but the organizing principle does not: higher-order function is controlled by the support of a tracked spectral branch, and when that support is concentrated on a small simplex set, a small targeted removal is sufficient to destroy the mode. The empirical graphs thus reproduce the same hidden higher-order vulnerability seen in the stylized bridge complex, establishing the mechanism as generic rather than construction-specific.

\subsection{Edge-Space Dynamics and Recovery Timescale}

We test whether the tracked eigenvalue has direct dynamical significance. On a Watts-Strogatz (WS) realization from the same benchmark family, we simulate edge-space relaxation under
\begin{equation}
    \dot{\bm{y}}=-L_1 \bm{y},
\end{equation}
using an initial condition dominated by the intact tracked mode with a small generic admixture. Harmonic components are projected out before normalization, ensuring that the late-time response is governed by the softest surviving nonharmonic branch.

WS networks are chosen for their combination of local clustering and sparse long-range connections, which generate slow, well-separated edge-space modes. This ensures that the tracked mode dominates the late-time relaxation, allowing a clear and reproducible measurement of the recovery timescale $t_{10^{-2}}$. While Erd\H{o}s-R\'enyi or spatial random networks could also be used, their slow modes are typically more distributed or less distinct, making branch-consistent verification less straightforward. WS networks therefore provide a representative and interpretable testbed without compromising generality.

For each attack state, we measure the threshold-crossing time
\begin{equation}
    t_{10^{-2}} = \inf \left\{t:\|\bm{y}(t)\|_2^2 \le 10^{-2}\right\},
\end{equation}
and compare it with the tracked-branch asymptotic prediction
\begin{equation}
    t_{10^{-2}}^{\mathrm{th}}
    = \frac{1}{2\widetilde{\mu}_1(f)}
    \log\!\left(\frac{|\widetilde{u}(f)^\top \bm{y}(0)|^2}{10^{-2}}\right),
\end{equation}
following the late-time single-mode approximation derived in Sec.~\ref{sec:Def}A.

\begin{figure}[t]
    \centering
    \includegraphics[width=1\linewidth]{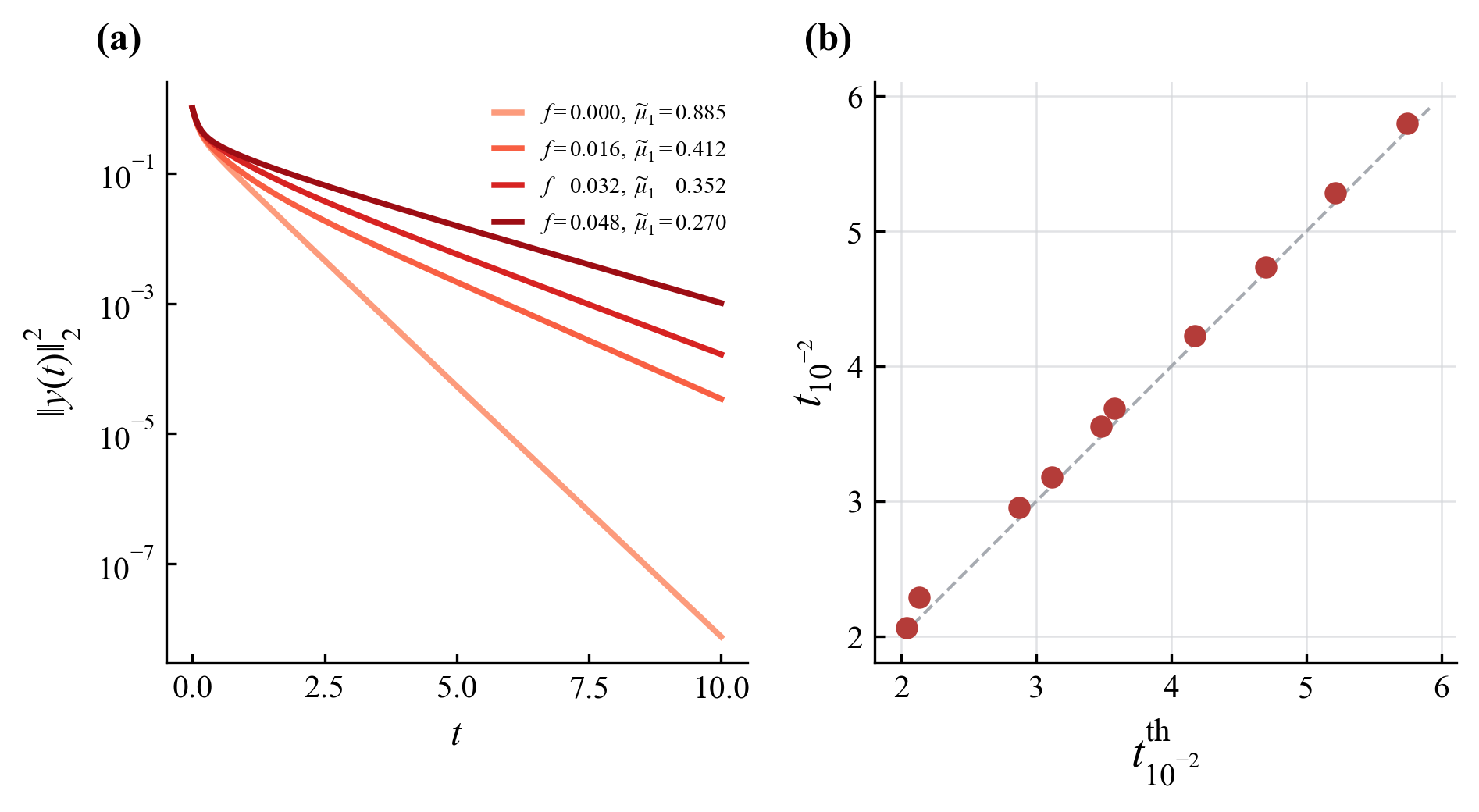}
    \caption{Dynamical validation of the tracked robustness measure. (a) Four representative edge-space relaxation curves under a common initialization dominated by the intact tracked mode on a Watts-Strogatz realization. Slower decay is observed for smaller tracked eigenvalues. (b) Measured energy-threshold times $t_{10^{-2}}$ versus the tracked-mode prediction $t_{10^{-2}}^{\mathrm{th}}$ across ten sampled pre-collapse attack states. The dashed line indicates perfect agreement; deviations reflect contributions from non-pure initial modes.}
    \label{fig:dynamics}
\end{figure}

The results demonstrate that the tracked branch accurately sets the dominant recovery timescale. Smaller $\widetilde{\mu}_1(f)$ values produce slower decay, leaving the energy above the threshold for longer times, as seen in Fig.~\ref{fig:dynamics}(a). Across ten sampled pre-collapse attack states, the measured $t_{10^{-2}}$ closely follows the single-mode prediction $t_{10^{-2}}^{\mathrm{th}}$ [Fig.~\ref{fig:dynamics}(b)]. Minor deviations arise because $t_{10^{-2}}$ is extracted from the total edge-space energy rather than from a pure single-mode amplitude, yet the systematic trend is clear: weakening the tracked branch consistently delays relaxation.

These observations indicate that the tracked branch is not only spectrally well-defined, but physically corresponds to the softest edge-space mode controlling late-time recovery. In other words, branch-consistent robustness directly maps onto functional dynamics, establishing a quantitative link between the spectral observable and the timescale of high-order functional collapse. The same mechanism is expected to generalize beyond Watts-Strogatz ensembles to other benchmark networks where the softest nonharmonic branch dominates relaxation.

\section{Discussion}

The central point of this work is that higher-order functional robustness is primarily a definitional problem. Under simplex deletion, the commonly used instantaneous smallest positive eigenvalue does not necessarily track a fixed physical degree of freedom. As the harmonic subspace expands, spectral reindexing induces branch switching, so the observable may refer to different nonharmonic modes at different stages of the process. In that situation, apparent robustness changes cannot be unambiguously interpreted as weakening of a single functional channel.

Fixing the spectral branch resolves this ambiguity at the level of the observable. By anchoring the first nonharmonic mode of the intact complex and following the same index throughout the dismantling trajectory, the robustness measure acquires a clear physical meaning as the decay of a specific edge-space response channel. Within this branch-consistent formulation, the simplex sensitivity is not an independent construction but the first-order perturbative response of the tracked mode. The resulting ranking therefore inherits its interpretation directly from the observable, rather than introducing an additional heuristic layer.

This corrected definition also clarifies the mechanism underlying rapid functional collapse. Because triangle deletion leaves the $1$-skeleton strictly unchanged, all graph-level observables remain invariant by construction. The observed collapse is therefore not related to connectivity degradation or edge removal, but is entirely generated by the coface contribution in the Hodge $1$-Laplacian. The separation between functional and structural robustness is thus exact in the present setting, and shows that pairwise descriptions can remain unchanged even when the dominant higher-order mode has already vanished.

The localization results further identify the structural origin of this vulnerability. The simplices that control the decay of the tracked mode are not those in the densest triangle-rich regions, but those carrying the largest projection of the mode itself. In many cases, this support is concentrated along sparse bridge-like carriers that mediate edge-space flow between mesoscopic regions. This explains why density-based and graph-based heuristics systematically underperform: they target geometric prominence, whereas the relevant object is modal support. The rapid collapse observed in the dismantling curves follows directly from this concentration, since a small subset of simplices can capture a large fraction of the total sensitivity mass.

The same support-driven mechanism is not restricted to stylized constructions. The synthetic benchmarks demonstrate that it persists across undirected, directed, and weighted clique complexes, indicating that it does not rely on a specific symmetry or weight distribution. The empirical networks further show that the effect survives in heterogeneous data, although the degree of localization varies across systems. What changes from case to case is the strength of concentration of the tracked mode, not the organizing principle itself.

The dynamical results provide an additional consistency check. The tracked eigenvalue governs the late-time decay of edge-space dynamics, setting the dominant recovery timescale of the system. This confirms that the branch-consistent observable captures a physically meaningful soft mode rather than a purely spectral surrogate. The agreement between measured relaxation times and the tracked-mode prediction shows that the spectral construction is directly connected to observable dynamical behavior.

\begin{figure}[t]
    \centering
    \includegraphics[width=1\linewidth]{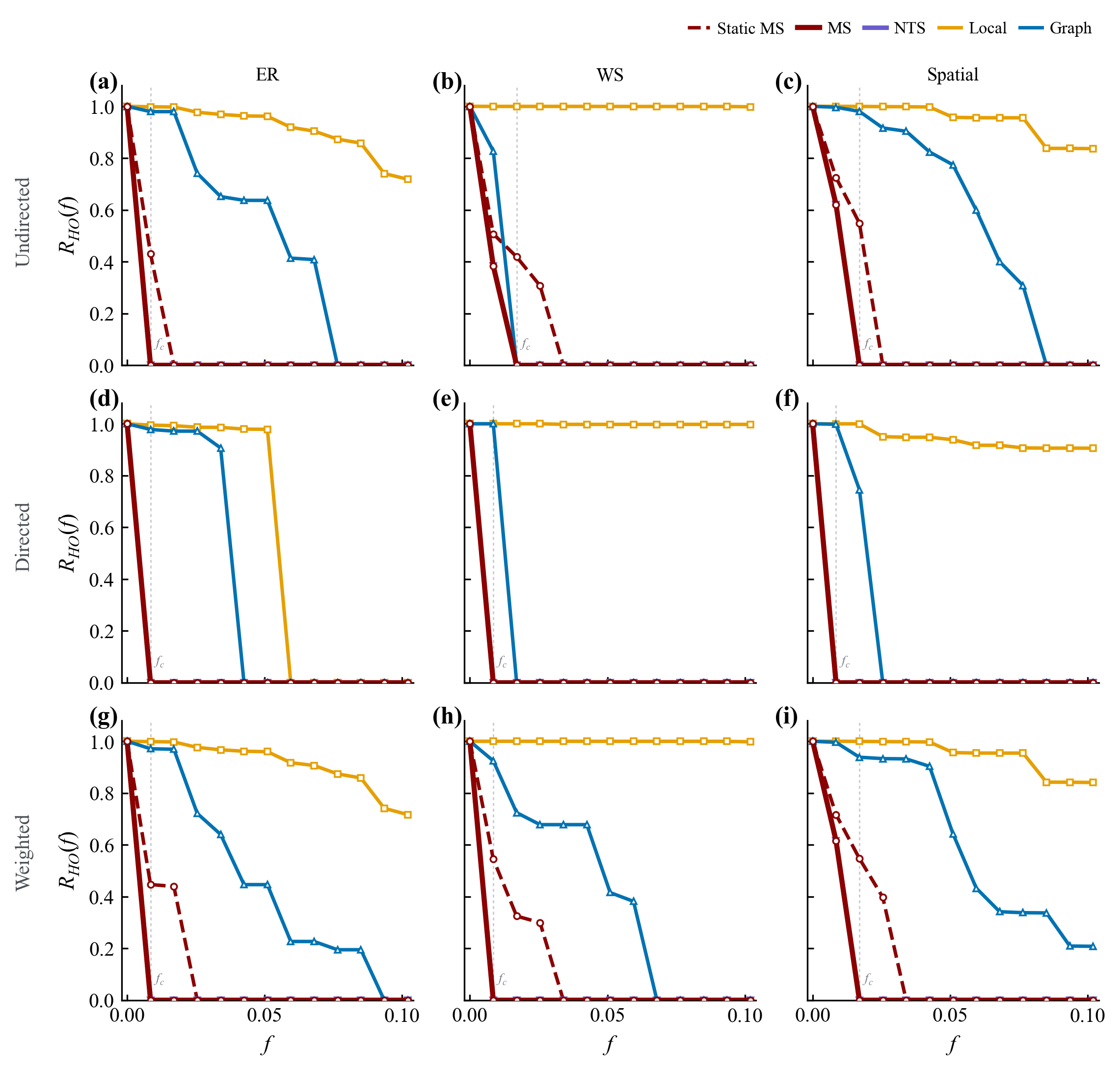}
    \caption{Step-by-step adaptive update of the tracked mode compared with static MS across synthetic ER, WS, and spatial ensembles. The dashed lines indicate static MS, which are consistent with the collapse curves shown in Fig.~\ref{fig:synthetic}. Solid lines correspond to adaptive step-by-step updating. The curves demonstrate that static MS already captures the collapse mechanism of the tracked higher-order mode, and adaptive updates serve only as a consistency check.}
    \label{fig:placeholder}
\end{figure}

The present formulation remains deliberately minimal. The numerical analysis focuses on triangle-based clique complexes and on a static ranking extracted from the intact tracked mode. To verify that static MS captures the same mechanism as an adaptive step-by-step update, we performed a small benchmark across ER, WS, and spatial ensembles, re-evaluating the tracked eigenvector after each triangle removal. The results, shown in Fig.~\ref{fig:placeholder}, confirm that the adaptive update produces qualitatively identical collapse curves, with only minor quantitative differences. This demonstrates that static MS already captures the underlying mechanism, while adaptive updating provides a consistency check rather than a new phenomenon.

In more general settings, including near-degenerate spectra, repeated near-crossings, or strongly asymmetric operators, one expects subspace-tracking or adaptive updates to become the natural continuation of the tracked mode. Extensions to higher-dimensional simplices or non-clique constructions are also straightforward. These generalizations, however, do not alter the central conclusion: higher-order robustness must be defined on a fixed spectral branch in order to represent the fate of a well-defined functional channel.

\section{Conclusion}

We have shown that the standard spectral observable commonly used in higher-order dismantling is generically ill-defined: the instantaneous smallest positive eigenvalue does not preserve the identity of the monitored mode under simplex deletion. This ambiguity is definitional rather than algorithmic; without fixing a spectral branch, robustness measurements do not track a single physical channel.

By anchoring the first nonharmonic branch of the intact complex, we define a branch-consistent higher-order functional robustness. Within this framework, Mode Sensitivity arises naturally as the first-order perturbative response of the tracked eigenvalue, making the associated dismantling strategy a direct consequence of the observable itself. This construction unifies robustness, sensitivity, and targeted removal at the level of a single spectral mode.

Numerical results reveal a hidden higher-order vulnerability: across synthetic and empirical clique complexes, removing only a small fraction of triangles suffices to extinguish the tracked mode, while the graph structure and all pairwise robustness indicators remain unchanged. This demonstrates that higher-order function constitutes a distinct robustness target, inaccessible from graph-level information alone.

A benchmark comparing static MS with step-by-step adaptive updates confirms that the collapse mechanism is insensitive to the update strategy. The results show that the underlying vulnerability is fully captured by the branch-consistent observable.

More broadly, these findings establish a simple principle: higher-order functional collapse is governed by the support of a tracked spectral branch in simplex space. When this support is strongly localized, a small targeted intervention can destroy the corresponding functional channel. Tracking such modes provides a compact, physically grounded framework for analyzing vulnerability in simplicial systems, validated across undirected, directed, and weighted settings, and naturally extendable to adaptive and higher-dimensional cases.
\bibliographystyle{apsrev4-2}
\bibliography{refs}

\end{document}